\documentclass{epl}
\usepackage{epsfig,amssymb,amsfonts,amsmath}
\usepackage{bm}
\usepackage{graphicx}
\usepackage{amsfonts}
\usepackage{amsmath}
\usepackage{amssymb}

\title{Dynamic fluctuations of elastic
lines in random environments}
\shorttitle{Elastic lines fluctuations}

\author{
Sebastian Bustingorry,
\inst{1}
Jos\'e Luis Iguain,
\inst{2}
Claudio Chamon,
\inst{3}
Leticia F. Cugliandolo,
\inst{4,5}
and 
Daniel Dom\'{\i}nguez
\inst{1}
}
\shortauthor{S. Bustingorry {\it et al}}
\institute{
\inst{1}Centro At{\'{o}}mico Bariloche, 
8400 San Carlos de Bariloche,
R\'{\i}o Negro, Argentina
\\
\inst{2}Departamento de F\'{\i}sica, FCEyN, 
Universidad Nacional de Mar del Plata, De\'an Funes 3350, 7600, 
Mar del Plata, Argentina
\\
\inst{3}Physics Department, Boston University, 590 Commonwealth Av., Boston 
MA 02215, USA
\\
\inst{4}Universit\'e Pierre et Marie Curie -- Paris VI, LPTHE UMR 7589,
4 Place Jussieu,  75252 Paris Cedex 05, France
\\
\inst{5}Laboratoire de Physique Th{\'e}orique de l'{\'E}cole Normale
Sup{\'e}rieure, 24 rue Lhomond, 75231 Paris Cedex 05, France
}

\date{\today}

\begin{document}
\maketitle

\begin{abstract} 
We study the fluctuations of the two-time dependent global roughness of
finite size 
elastic lines in a quenched random environment.  We propose a scaling
form for the roughness distribution function that accounts for the
two-time, temperature, and size dependence. At high temperature and in
the final stationary regime before saturation the fluctuations are as
the ones of the Edwards-Wilkinson interface evolving from typical
initial conditions.  We analyze the variation of the scaling function
within the aging regime and with the distance from saturation. We
speculate on the relevance of our results to describe the fluctuations
of other non-equilibrium systems such as models at criticality.
\end{abstract}

The study of dynamic fluctuations may serve to grasp the nature of
non-equilibrium phenomena.  Building upon previous analysis of critical
static and dynamic fluctuations~\cite{static-inter,Antal},
R\'acz~\cite{Racz} proposed to use the scaling functions
characterizing the fluctuations of macroscopic - global - observables
in {\it non-equilibrium steady states} as a classification tool. 
By introducing `universality classes' in this way, one could then use 
them to uncover symmetries and dynamic mechanisms in experimental 
systems. 

 Another class of non-equilibrium phenomena, including important
problems such as glassiness and coarsening, is characterized by a 
slow relaxation with loss of stationarity~\cite{reviews,Leto}. It is by
now becoming clear that the study of fluctuations~\cite{exps} 
is necessary to understand the mechanism for the slowing down and the 
{\it aging non-equilibrium relaxation} in these cases
too~\cite{Chamon-etal,Chamon-LFC-Yoshino}.

An elastic line under the effect of quenched disorder is a relatively 
simple system with
many aspects of glassiness due to the competition between elasticity
and disorder. Its physical realizations are manifold,
including interfaces in random ferromagnets~\cite{HH}, crack
propagation~\cite{crack}, and vortex lines in high-$T_c$ dirty
superconductors~\cite{fisher}.  The global noise and disorder averaged
displacement and linear response age with a {\it
multiplicative} temporal scaling~\cite{Yosh98,us} similar to what is
found at criticality~\cite{Zannetti} and in Sinai
diffusion~\cite{Pierre}.  In this Letter we study the
line's fluctuating relaxation by analyzing the probability
distribution functions (pdfs) of the two-time dependent global roughness
and we discuss it in comparison to what has been found in other aging
systems~\cite{exps,Chamon-etal,Chamon-LFC-Yoshino} and the
Edwards-Wilkinson (EW) interface~\cite{Antal}.


We study a lattice string of length $L$ directed along the $y$
direction of a rectangular square lattice of transverse size
$M\gg L^{2/3}$ $(M=10^4 , L=500)$ ensuring the existence of many
nearly equivalent ground states.  The line segments, $x_y$
($y=1,\cdots,L$), obey the restricted solid-on-solid (SOS) rule
$|x_y-x_{y-1}|=0, 1$.  A quenched random potential $V$ taking
independent values on each lattice site is drawn from a uniform
distribution in $[-1,1]$; we use $10^5-10^7$ realization depending on 
$L$. The initial configuration ($t=0$) is drawn from 
the equilibrium distribution at the 
high $T=5$. At each microscopic time step we attempt a
move of a randomly chosen segment to one of its neighbours restricted
by the SOS condition and we accept it with the heat-bath
rule.  One Monte Carlo (MS) step is defined as $L$
update attempts. In the following we use adimensional time, space and
energy scales~\cite{Yosh98}.

The glassy phenomenon appears in this model as a dynamic
crossover~\cite{Yosh98,us}.  For all observation times, $t_{obs}$,
that are longer than a size and temperature dependent {\it
equilibration time}, $t_{eq}$, the dynamics is
stationary. Instead, for $t_{obs} < t_{eq}$ the relaxation occurs out
of equilibrium as demonstrated by two-time correlations that age. The
latter are measured as follows. After equilibration at high $T$ the
system is quenched to low $T$ and time is set to zero. The line
relaxes until a waiting-time, $t_w$, when the quantities of interest
are recorded and later compared to their values at a subsequent time
$t$.

In this Letter we focus on the {\it two-time} global roughness 
\begin{equation} 
w^2(t,t_w) \equiv L^{-1} 
\textstyle{\sum_{y=1}^L} |\delta x_y(t)-\delta x_y(t_w)|^2 
\; , 
\label{def:roughness}
\end{equation} 
with $\delta x_y(t) \equiv x_y(t) - \overline x(t)$ the 
distance from the center of mass, 
$\overline{x}(t)\equiv L^{-1}\sum_{y=1}^L x_y(t)$.
Before any averaging, $w^2$ fluctuates when changing 
the disorder and thermal noise realizations.

At high $T$, neglecting a short transient, the disorder
and thermal averaged roughness, $\langle w^2\rangle$, is stationary and
crosses over from growth to saturation at $t_x\sim
L^{z}$~\cite{Barabasi-Stanley}:
\begin{equation}
\langle w^2(t,t_w) \rangle \sim L^{\zeta} \; f(\Delta t/t_x) 
\; , 
\qquad 
\Delta t \equiv t-t_w
\; ,
\label{eq:tau}
\end{equation}
where the scaling function obeys $f(x) \sim x^\beta$ for $x \ll 1$
and $f(x) \sim 1$ for $x \gg 1$, with $\zeta$, $\beta$ and $z=\zeta
/ \beta$ the roughness, growth and dynamic exponents,
respectively. For the EW line in $1+1$ dimensions $\zeta=1$,
$\beta=1/2$ and $z=2$. In the presence 
of disorder, $\zeta$ is expected to take a `thermal' value,
$\zeta_{th}$, for $L<L_c(T)$, and a larger `disorder' dominated value,
$\zeta_{dis}$, for $L>L_c(T)$, both being
$T$-independent~\cite{Barabasi-Stanley}. 

\begin{figure}[!tbp]
\centerline{
\includegraphics[angle=-0,width=11cm,clip=true]{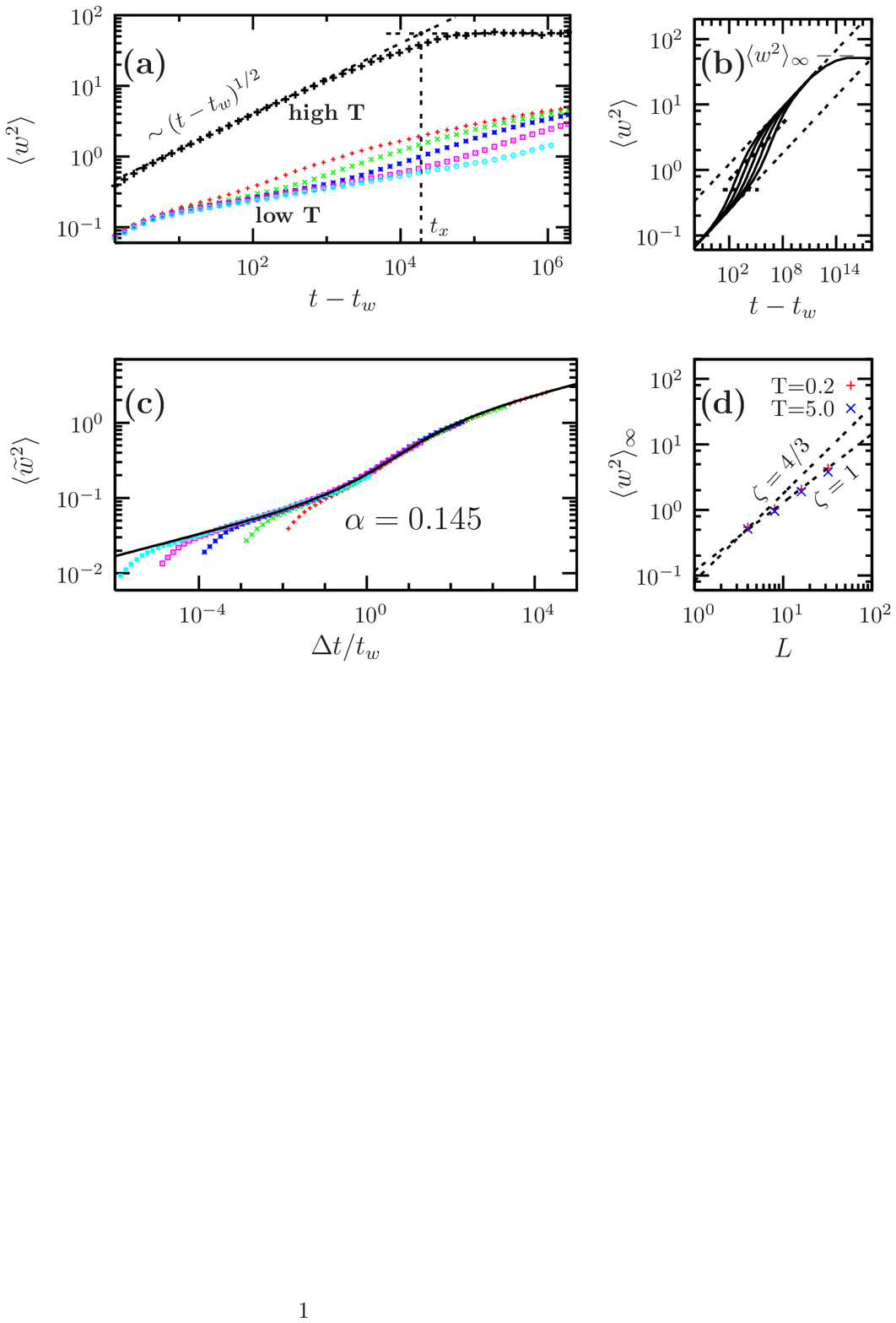}
}
\caption{(Colour on-line.)
Panel (a): thermal and disorder averaged square roughness as
a function of time-difference for a line with $L=500$. The top curve
is for $T=5$; the dotted lines indicate the sub-diffusive regime,
saturation and the characteristic time $t_x$. The high-$T$ dynamics is
stationary and conforms with the scaling in (\ref{eq:tau}). The bottom
curves are for $T=0.2$, and $t_w=10^2, \, 10^3, \, 10^4, \, 10^5, \,
10^6$ MCs from top to bottom.  At low $T$, stationarity is lost after
an initial $\Delta t$ regime.  Saturation is out of the $\Delta t$
window for all low $T$ curves.  Panel (b): sketch of the
time-difference dependence of the averaged roughness for several
waiting-times. The solid lines have been drawn using the function
${\cal G}$ in eq.~(\ref{eq:fitting}) and a crossover to saturation at
$\langle w^2\rangle_\infty$. The dashed lines represent the asymptotic
power laws $c_{1,2}(T) \Delta t^\alpha$.  The horizontal dotted line
indicates a path of constant $\langle w^2\rangle/\langle
w^2\rangle_\infty$.  The intersection of the curves with the
inclined dotted line correspond to constant $\nu_1$, see
eq.~(\ref{eq:def-nu1}).  Panel~(c): scaling of the data in panel (a)
using eqs.~(\ref{eq:fitting0})-(\ref{eq:fitting}); the solid line is
${\cal G}$ in (\ref{eq:fitting}) with $\alpha=0.145$, $A=0.28$,
$B=0.35$, $C=0.56$ and $D=5.01$. 
Panel (d): $L$-dependence of the
saturation value, $\langle w^2\rangle_\infty$, at high and low $T$.}
\label{fig1}
\end{figure}

For not too short $L$ and sufficiently low $T$, $t_{eq}$ goes
beyond the numerically accessible time-window and $\langle w^2\rangle$
depends on $t_w$ explicitly. Before saturation 
one expects
\begin{eqnarray}
&& 
\langle w^2(t,t_w) \rangle 
\sim 
\ell^\zeta(t_w) \; \langle \widetilde{w}^2[\ell(t)/\ell(t_w)]\rangle
\equiv
\ell^\zeta(t_w) \; {\cal F}[\ell(t)/\ell(t_w)]
\label{eq:fitting0}
\end{eqnarray} 
with $\ell(t)$ a growing length (dimensions are restored by prefactors
that we omit).  This form approaches a stationary regime when $t_w \ll
\Delta t\ll t_x$ if ${\cal F}(x) \sim x^{\zeta}$ for $x\gg 1$, and
complete saturation at $L^\zeta$ when $\ell(t) \to L$.  We found data
collapse using $\ell(t)\sim t^{\alpha/\zeta}$ with a {\it small} exponent
\cite{footnote} and ${\cal F}[\ell(t)/\ell(t_w)]={\cal G}(\Delta
t/t_w)$ with
\begin{equation}
{\cal G}(x) \sim  x^\alpha A\, 10^{B g(x)}
\to x^\alpha A\, 10^{\pm B} 
\;,
\label{eq:fitting}
\end{equation}
in the limits $x\gg 1$ and $x\ll 1$, respectively, 
and
$g(x)=\tanh[C \log_{10}(x/D)]$ which 
does not have a special significance but serves
to select the working times $\Delta t$ and $t_w$ below. 
Note that on
the two asymptotes stationarity is recovered:
\begin{equation}
\langle w^2\rangle \sim  c_{1,2}(T) \, \Delta t^\alpha
\; ,
\;\;\;\;  
\mbox{with}
\;\;
\alpha < \beta_{EW} = 0.5
\; ,
\label{eq:straight}
\end{equation}
and different proportionality constants 
$c_{1,2}=A\, 10^{\mp B}$.
$\alpha$ is then a generalization of the growth exponent, $\beta$
and $\alpha< \beta_{EW}$ reflects that disorder slows down the 
dynamics. 
In Fig.~\ref{fig1} we show $\langle w^2\rangle$ at high and low $T$:
panels (a) and (c) display numerical data and panel (b) presents a
sketch of the size and time-dependence of $\langle w^2\rangle$ at low
$T$. Details are given in the caption.  

Even though $\langle w^2\rangle$ weakly depends on the polymer size before
saturation, the saturation value strongly depends on $L$
and it turns out to be important to describe the fluctuations.
Transfer matrix calculations show that the static roughness $W^2_y
\equiv \langle [x_y-x_0]^2\rangle$ with $x_0$ fixed crosses over
from $W^2_y \sim y$ to $W^2_y \sim y^{4/3}$ at a length $y^*\sim 5$  
at $T=0.2$~\cite{Yosh98}. One might expect a similar
crossover for the asymptotic global roughness (\ref{def:roughness})
though the crossover length $y^*$ should not necessarily be the same.  With
MC simulations we can only determine $\langle w^2\rangle_\infty \equiv
\lim_{\Delta t \gg t_x} \langle w^2 \rangle$ for rather short polymer
lengths ($L\leq 32$) both at high and low temperature. 
Figure~\ref{fig1}(d) shows its $L$-dependence at $T=5$ and $T=0.2$, 
two values that we shall use in the rest of the
Letter. The dependence is linear with no sign of the crossover to the
disorder-dominated regime up to the maximum length, $L=50$.

\begin{figure}[!tbp]
\centerline{
\includegraphics[angle=-0,width=9cm,clip=true]{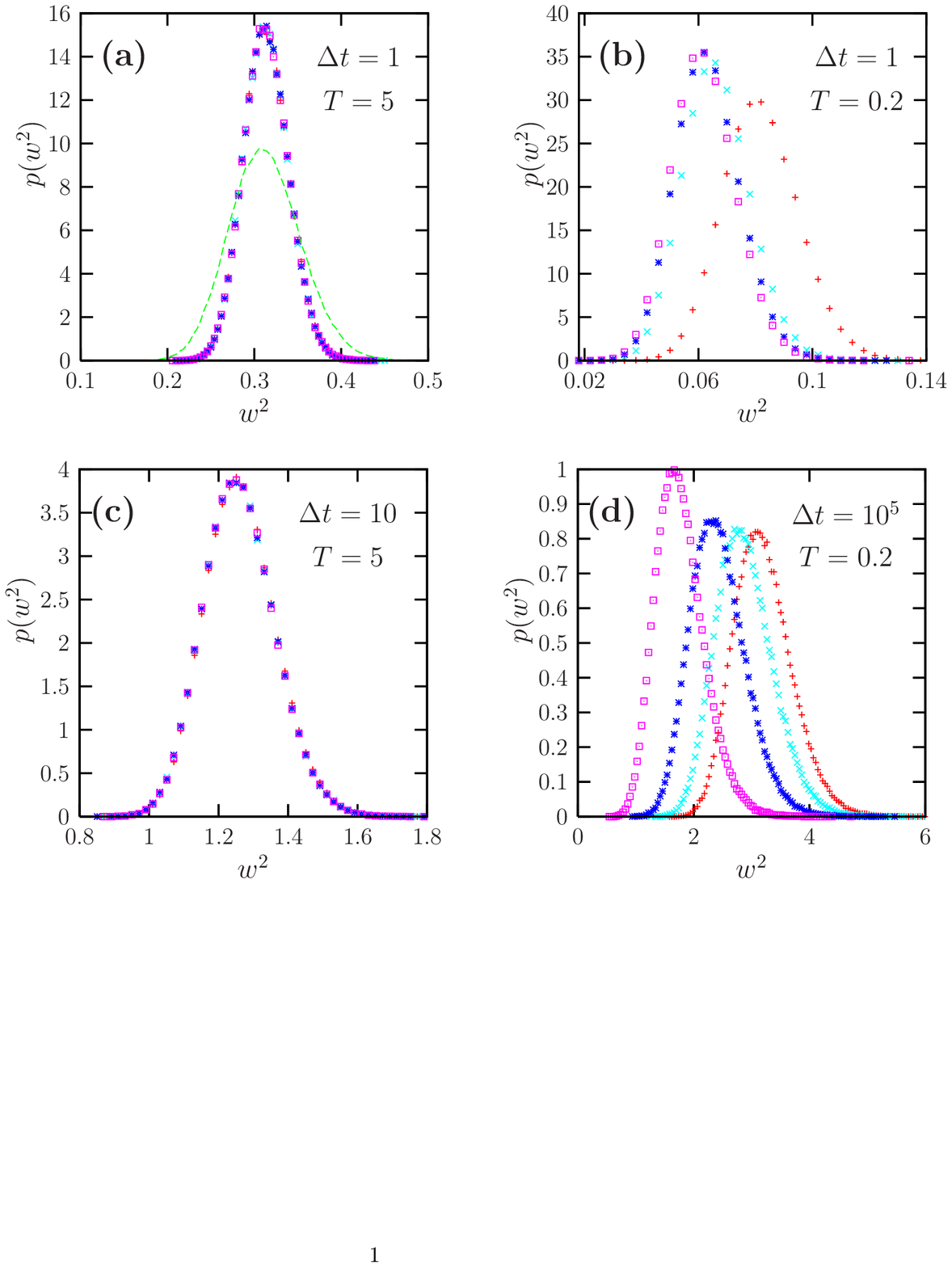}
}
\caption{(Colour on-line.)
Bare square roughness fluctuations at $T=5$ [(a) and (c)] and
$T=0.2$ [(b) and (d)] for a line with $L=500$.  The time-differences
are given in the key.  The waiting-times are $t_w=10^1, \, 10^2, \,
10^3, \, 10^4$ MCs [from right to left in panels (b) and (d)].  In panel (a)
one curve (dashed green) for $L=100$ and the same parameters is
included to highlight the $L$-dependence of the fluctuations. }
\label{fig2}
\end{figure}

We now turn to the study of fluctuations.  In Fig.~\ref{fig2} we show
the pdf of square roughness fluctuations, $p(w^2)$, for a line with
$L=500$ at $T=5$ [left, panels (a) and (c)] and $T=0.2$ [right,
panels (b) and (d)] for several values of $t_w$ and $\Delta t$ given
in the keys.  At $T=5$ there is no $t_w$ dependence. The
pdfs evolve with $\Delta t$ and they basically take
the EW form~\cite{Antal}. At low $T$ a clear
$t_w$ dependence appears at sufficiently (but not too) long $\Delta t$'s.

In order to understand $p(w^2)$ we propose a scaling form that we
later put to the test numerically.  In the low $T$ regime, for times
such that equilibration has not been reached ($t_w \ll t_{eq}$)
$p(w^2)$ depends on the (already adimensional) times $t$ and $t_w$,
the system size $L$ and temperature $T$. With a simple change of
variables we measure $w^2$ in units of its average $\langle w^2
\rangle$: $p(w^2;t,t_w,L,T) = \langle w^2 \rangle^{-1}
\rho(w^2/\langle w^2\rangle;t,t_w,L,T) $.  Using
eq.~(\ref{eq:fitting0}), and the monotonicity of $\ell$ and 
${\cal G}$, the dependence on $t$ and
$t_w$ can be replaced by a dependence on $\langle w^2\rangle$ and
$\langle \widetilde w^2\rangle$. $L$ can be traded for the
$T$-dependent saturation value $\langle w^2\rangle_\infty$.  Now, our
{\it scaling hypothesis} is that the dependence on $\langle
w^2\rangle$ occurs only in comparison with the saturation
value:
\begin{equation}
\langle w^2 \rangle p(w^2) = 
\overline\Phi\left(w^2/\langle w^2\rangle;
\langle \widetilde w^2\rangle,
\langle w^2\rangle/\langle w^2\rangle_\infty,T
\right)
\; . 
\label{eq:hypothesis-mult}
\end{equation}
This form holds exactly at high temperatures and for the EW
interface. In these cases disorder is irrelevant or even absent and
the $t_w$-dependence, and hence the $\langle \widetilde
w^2\rangle$-dependence, disappears.  In addition, all $T$ dependence
enters through $\langle w^2\rangle_\infty$ and one has $\langle
w^2\rangle p(w^2) = \Phi_{EW}(w^2/\langle w^2\rangle, \langle
w^2\rangle/\langle w^2\rangle_\infty)$~\cite{Antal}.

To account for the $\langle \widetilde w^2\rangle$ dependence
we introduce the parameter
\begin{equation}
\nu_1 \equiv
[\langle \widetilde w^2\rangle -c_1 (\Delta t/t_w)^\alpha]/
[(c_2-c_1) (\Delta t/t_w)^\alpha]
\label{eq:def-nu1}
\end{equation}
within the description of the averaged data with 
the power fit $\ell(t) \sim t^{\alpha/\zeta}$. $\nu_1$ 
varies between $0$ and $1$ when $\langle \widetilde w^2\rangle$ 
moves from the first to the second asymptotic power laws before saturation,
see the dashed lines in Fig.~\ref{fig1}(b) and eq.~(\ref{eq:straight}).
We call $\nu_2\equiv 
\langle w^2\rangle/\langle w^2\rangle_\infty$ the second 
parameter in (\ref{eq:hypothesis-mult}).
Then, by choosing groups of $(L,\Delta t,t_w)$ we vary $\nu_1$ and 
$\nu_2$. 
In the following we test 
the following restatement of our scaling hypothesis 
\begin{equation}
\langle w^2 \rangle p(w^2) = 
\Phi\left(w^2/\langle w^2\rangle;\nu_1,\nu_2,T\right)
\; . 
\label{eq:hypothesis-mult1}
\end{equation}

\begin{figure}[!tbp]
\centerline{
\includegraphics[angle=-0,width=11cm,clip=true]{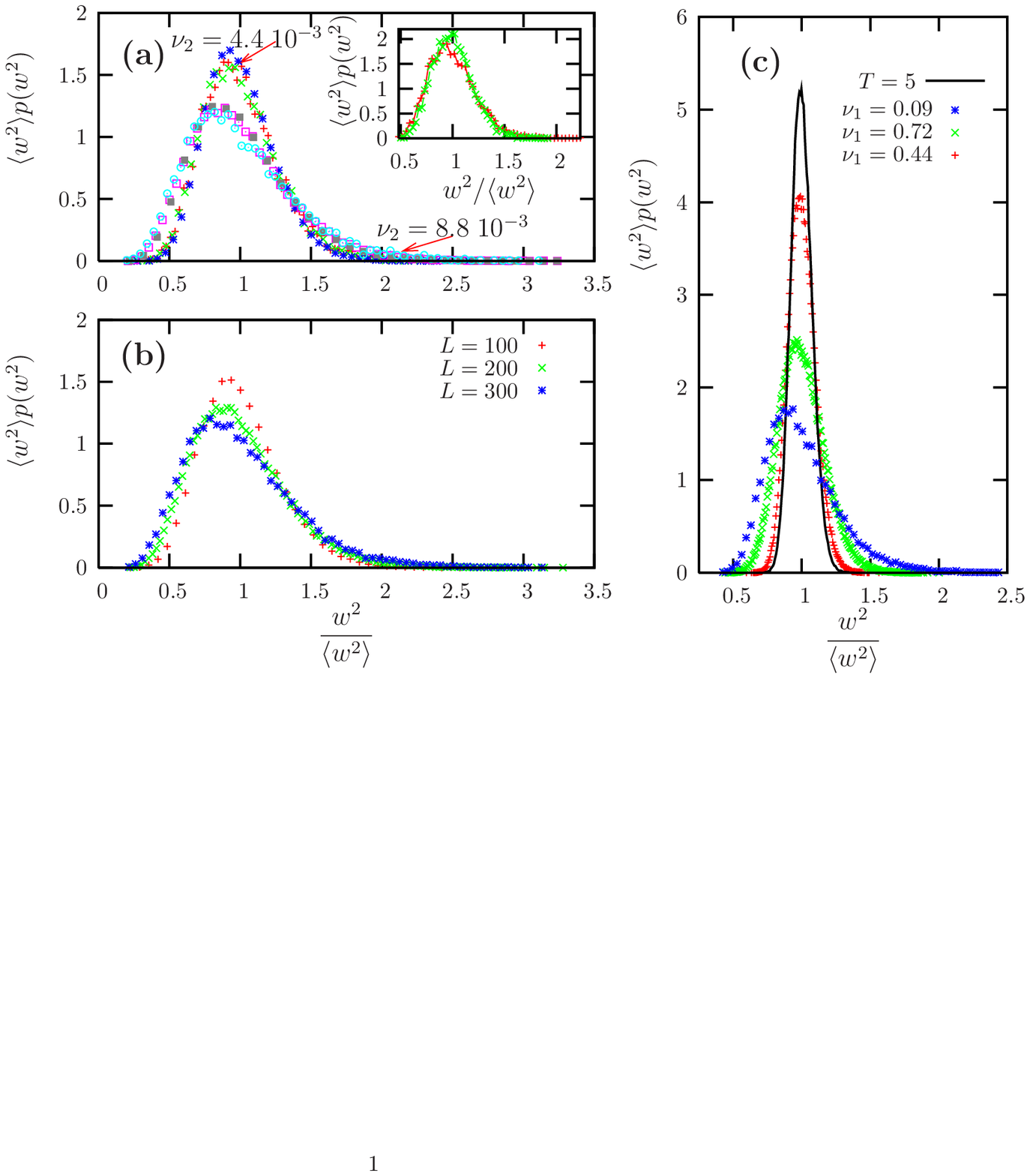}
}
\caption{(Colour on-line.)  Panel~(a): check of the scaling
  (\ref{eq:hypothesis-mult}) using $\langle w^2\rangle_\infty \propto
  L$ ($\zeta=1$).  The gray filled squares ($L=100$, $t_w=100$), pink
  open squares ($L=200$, $t_w=1.19\, 10^4$) and cyan open circles
  ($L=300$, $t_w=1.95\, 10^5$) data are at the same distance from
  saturation ($\nu_2=\langle w^2\rangle/L = 8.8\;10^{-3}$), and have
  the same `aging' roughness ($\nu_1=0.389$, $\Delta t/t_w=10$).  The
  blue $\star$ ($L=200$, $t_w=10^2$), green $\times$ ($L=400$,
  $t_w=1.19\, 10^2$) and red $+$ ($L=600$, $t_w=1.95\, 10^5$) data are
  for $\nu_2=4.4\;10^{-3}$ and the same $\nu_1$ as the other
  group. Inset: the pdf for the continuous model in~\cite{us}. Panel
  (b): test of the scaling (\ref{eq:hypothesis-mult}) using $\langle
  w^2\rangle_\infty \propto L^{4/3}$ ($\zeta=4/3$) and 
  $\nu_2 = \langle
  w^2\rangle/L^{4/3}=1.3\;10^{-3}$. Panel~(c): Dependence on
  $\nu_1$ and comparison with the high-$T$ behaviour (black solid
  line).  One gets deeper into the aging regime in the order blue
  $\star$ ($\Delta t=2.25~10^5, t_w=10^6, \nu_1=0.09$), green $\times$
  ($\Delta t=1.50~10^3, t_w=10^2, \nu_1=0.44$), red $+$ ($\Delta
  t=1.50~10^2, t_w=10^0,\nu_1=0.72$). $\nu_2 = \langle
  w^2\rangle/L=2\;10^{-3}$.}
\label{fig3}
\end{figure}

In Fig.~\ref{fig3}(a) we plot data for 
$\nu_1=0.389$ and different $\nu_2$.
We use $\langle w^2\rangle_\infty\sim L$ and 
we plot six sets of data corresponding to two values of $\nu_2=
\langle w^2\rangle /L$ (the $T$-dependent prefactor is irrelevant since 
we are working at constant $T$).
The two sets of data with the same $\nu_2$ collapse rather well. 
The dependence on $\nu_2$ is similar to the one found for the EW line:
for small $\nu_2$ the scaling function has an approximately log-normal
form, it is positively skewed and it gets wider with increasing
$\nu_2$. In panel (b) we use instead $\langle w^2\rangle_\infty\sim
L^{4/3}$ and the scaling variable $\nu_2= \langle w^2\rangle
/L^{4/3}$. The collapse is now lost in accordance with our hypothesis
and the fact that $\langle w^2\rangle_\infty \sim L$ for the lengths
used.  (It would of course be desirable to have an independent
determination of $\langle w^2 \rangle_\infty$ to confirm our
conclusions.)
 
In Fig.~\ref{fig3}(c) we analyze the evolution of $\Phi$ with $\nu_1$.
$\Phi$ resembles the high $T$ result with the same distance from
saturation (black solid line) when $\nu_1 \to 1$ and 
$\langle w^2\rangle \sim c_2(T) \Delta t^\alpha$.
It progressively deviates from the high $T$ form for decreasing
$\nu_1$. Decreasing $\nu_1$ while keeping $\nu_2$ fixed has a similar
effect as increasing $\nu_2$ while keeping $\nu_1$ fixed: $\Phi$ gets
closer to the asymptotic disorder-free form, $\lim_{\Delta t\gg t_x}
\Phi_{EW}$. Moreover, at fixed and small $\nu_2$ the $\nu_1$-dependent
pdfs can also be quite well described with a log-normal function with
two $\nu_1$-dependent parameters.

The $\nu_1$ dependence can be rationalised from the point of view of
the dynamics in a free-energy landscape.  For each $t_w$ there is a
sufficiently long $\Delta t$ such that $\langle w^2\rangle
\sim c_2(T) \Delta t^\alpha$ and the waiting-time dependence
disappears (see Fig.~\ref{fig1}).  After these very long $\Delta t$'s
one allows the system to get out of the traps occupied at $t_w$,
sub-diffuse with $\alpha < \beta_{EW} = \frac12$ but fluctuate
similarly to the high $T$ and Gaussian cases once the good scaling
variables are chosen, as shown by the fact that
$p(w^2)$ gets very close to the continuous curve in
Fig.~\ref{fig3}(c).  In the opposite limit in which $\langle
w^2\rangle$ has not deviated much from the first asymptote
the system is quite trapped and has not had enough time-difference,
$\Delta t$, to explore a relevant part of the free-energy landscape.
The fluctuations are then very different from the high-$T$ and 
disorder-free ones.

In the inset to Fig.~\ref{fig3}(a) we show 
data from the Langevin dynamics of a 
{\it continuous} model of an elastic line in a $3d$ disordered
environment~\cite{zimanyi,us} that confirms the scaling form and the
generic trend.  
We shall present more details on the 
functional form of $\Phi$ in the lattice and continuous 
models as well as on the $T$-dependence  in a longer publication~\cite{long}.

Summarizing, we studied the effect of disorder on the low-$T$ averaged
and fluctuating two-time roughness of elastic lines.  On the one hand
disorder introduces aging at sufficiently low temperatures. Thus, a
time-delay that increases with the waiting-time is necessary to enter
a stationary regime next to saturation. On the other hand, it slows
down the evolution and the averaged roughness in this
last dynamic regime undergoes sub-diffusion with a smaller exponent
than in the EW case.  Well before saturation
the two-time roughness fluctuations can be represented by a scaling
function, $\Phi$, that depends on where in the aging regime the
dynamics take place (parametrized by $\langle \widetilde w^2\rangle$
or $\nu_1$) and the distance from saturation (given by $\nu_2\equiv
\langle w^2\rangle/ \langle w^2\rangle_\infty$).  At high $T$, and at
low $T$ but close to the sub-diffusive regime before saturation,
disorder is quite irrelevant and $\Phi$ approaches the EW form
$\Phi_{EW}$ depending only on $\nu_2$. At intermediate time-scales,
where the waiting-time dependence is explicit, $\Phi$ deviates from
the free result and becomes broader and less symmetric. The dependence
on the distance to saturation at fixed `aging' parameter $\nu_1$ is
similar to the one in the EW line.

We conjecture that this kind of scaling appears also in line problems
in different universality classes, {\it e.g.}  KPZ~\cite{Brazil},
and/or for other quantities such as the maximum height with respect to
the mean~\cite{Comtet}.  It would also be interesting to study these
features in the out of equilibrium dynamics of critical
systems~\cite{Zannetti} and random walkers in random environments~\cite{Pierre}
with a multiplicative scaling of two-time correlations and responses.

This study extends the proposal in \cite{Chamon-etal} for the
fluctuations in aging systems with additive scaling -- correlations
that approach a well-defined and finite plateau to later further decay
to zero -- to problems with {\it diffusive aging} and a strong
dependence on the system size. In the former cases one could explain
the scaling and generic form of the pdfs as a consequence of the
emergence of a symmetry, {\it time reparametrization invariance}, in
the long time dynamics.  The presence of the symmetry allowed one to
propose an effective action for the most important fluctuations, and
derive from it the pdfs of local correlations and responses that have
been partially confronted to experimental measurements in colloidal
suspensions~\cite{exps} and simulations of kinetically constrained
models and finite dimensional spin-glass models~\cite{Chamon-etal}.
The study of the symmetry properties of the long-times dynamics of
elastic lines in random environments and further identification of an
effective sigma-model-type action for the fluctuations remains to be
done. It should be possible to use it to
compute~(\ref{eq:hypothesis-mult}).

We acknowledge useful discussions with A. Kolton and 
H. Yoshino and financial
support from SECYT-ECOS P. A01E01 (LFC, DD), 
 CNEA, CONICET PIP05-5596 (SB, DD) and PIP05-5648 (JLI),
     ICTP-NET-61 (DD, JLI), ANPCYT PICT04-20075 (JLI),
Fundaci\'on Antorchas (SB), NSF-CNRS INT-0128922 (CC,
LFC), PICS 3172 (LFC), and NSF DMR-0305482 and DMR-0403997 (CC). 
SB and LFC thank the
UNMDP for hospitality.
JLI is indebted to the R\'eseau
Qu\'eb\'ecois de Calcul de Haute Performance 
for generous allocations of computer resources. LFC is a
member of IUF.


\begin{thebibliography}{}

\bibitem{static-inter}
G. Foltin {\it et al}, Phys. Rev. E {\bf 50}, R639 (1994).
M. Plischke {\it et al}, Phys. Rev. E {\bf 50}, 3589 (1994).
Z. R\'acz and M. Plischke, Phys. Rev. E {\bf 50}, 3530 (1994).
S. T. Bramwell {\it et al.} Nature {\bf 396}, 552 (1998).

\bibitem{Antal} T. Antal and Z. Racz, Phys. Rev. E {\bf 54}, 2256 (1996).

\bibitem{Racz} Z. R\'acz, SPIE proceedings Vol. 5112, 248 (2003). 

\bibitem{reviews} 
E. Vincent, J. Hammann, M. Ocio, J-P Bouchaud, and L. F.
Cugliandolo, in {\it Complex behaviour in glassy systems}
E. Rubi ed. (Springer-Verlag, 1997), cond-mat/9607224. 

\bibitem{Leto} L. F. Cugliandolo 
in {\it Slow Relaxations and Nonequilibrium Dynamics in
Condensed Matter}, J. -L. Barrat {\it et al.} Eds., (Springer, Berlin,
2002).

\bibitem{exps}  
R. Courtland, and E. Weeks, J. Phys. C 15, S359 (2003). 
K. S. Sinnathamby, H. Oukris, and N. E. Israeloff, cond-mat/0412378. 
L. Cipelletti and L. Ramos, J. Phys. C {\bf 17}, R253 (2005). 
A. Duri, H. Bissig, V. Trappe, and L. Cipelletti, 
Phys. Rev. E {\bf 72}, 051401 (2005).
P. Wang, C. Song, and H. Makse, Nature Physics {\bf 2}, 526 (2006).


\bibitem{Chamon-etal} C. Chamon, M. P. Kennett, H. Castillo, and 
L. F. Cugliandolo, Phys. Rev. Lett. {\bf 89}, 217201 (2002). 
H. E. Castillo, C. Chamon, L. F. Cugliandolo, and M. P. Kennett, 
Phys. Rev. Lett. {\bf 88}, 237201 (2002); 
H. E. Castillo, C. Chamon, L. F. Cugliandolo, J. L. Iguain, and M. P. Kennett, 
Phys. Rev. B {\bf 68}, 134442 (2003).  
C. Chamon, P. Charbonneau, L. F. Cugliandolo, D. Reichman, and M. Sellitto,
J. Chem. Phys. {\bf 121}, 10120 (2004).

\bibitem{Chamon-LFC-Yoshino} C. Chamon, L. F. Cugliandolo, and H. Yoshino, 
J. Stat. Mech (2006) P01006.

\bibitem{HH} D. A. Huse and C. L. Henley, 
Phys. Rev. Lett. {\bf 54}, 2708 (1985).

\bibitem{crack} A. Hansen and E. L. Hinrichsen, and S. Roux, 
Phys. Rev. Lett. {\bf 66}, 2476 (1991). 

\bibitem{fisher} See, {\it e.g.} G. Blatter, M. V. Feigel'man, V. B.
  Geshkenbein, A. I. Larkin, and V. M. Vinokur, Rev. Mod. Phys. {\bf
    66}, 1125 (1994); T. Nattermann and S. Scheidl, Adv. in Phys. {\bf
    49}, 607 (2000).

\bibitem{Yosh98}  
H. Yoshino, J. Phys. A {\bf 29}, 1421 (1996);
Phys. Rev. Lett. {\bf 81}, 1493 (1998); and unpublished.
A. Barrat, Phys. Rev. E {\bf 55},  5651 (1997).

\bibitem{us} S. Bustingorry, L. F. Cugliandolo, and D. Dom\'{\i}nguez, 
Phys. Rev. Lett. {\bf 96}, 027001 (2006).

\bibitem{Zannetti} A. Gambassi and P. Calabrese, J. Phys. A {\bf 38}, R133
  (2005) and references therein. 

\bibitem{Pierre} L. Laloux and P. Le Doussal, Phys. Rev. E 
{\bf 57}, 6296 (1998). P. Le Doussal, C. Monthus, and D. S. Fisher
Phys. Rev. E {\bf 59}, 4795 (1999).

\bibitem{Barabasi-Stanley} 
A-L Barab\'asi and H. E. Stanley, {\it Fractal
concepts in surface growth} (Cambridge University Press, Cambridge,
1995). T. Halpin-Healey and Y-C Zhang, Phys. Rep. {\bf 254}, 215 (1995).

\bibitem{footnote} 
Simulations of a similar model~\cite{Kolton} suggest that $\ell$
crosses over from power-law to logarithmic growth; it is very hard to
distinguish a logarithmic growth from a power law with a small
exponent as the one we use here.  Note that (\ref{eq:fitting0}) has 
a $\Delta t \ll t_w$ stationary regime only
if $\ell(t)\sim t^a$.

\bibitem{Kolton} A. Kolton, A. Rosso, and T. Giamarchi, 
Phys. Rev. Lett. {\bf 95}, 180604 (2005).


\bibitem{super} D. R. Nelson and H. S. 
Seung, Phys. Rev. B {\bf 39}, 9153 (1989).

\bibitem{zimanyi} 
C. Reichhardt, A. van Otterlo, and G. T. Zimanyi, Phys. Rev. Lett. 
{\bf 84}, 1994 (2000).



\bibitem{Brazil} F. D. A. Aar\~ao Reis, cond-mat/0508238. 

\bibitem{Comtet} S. Majumdar and A. Comtet, Phys. Rev. Lett. {\bf 92}, 225501 (2004).

\bibitem{long} S. Bustingorry {\it et al}, in preparation. 



\end{thebibliography}
\end{document}